\documentclass{PoS}

\usepackage{amsmath, amsfonts, amssymb}
\usepackage{graphicx}
\usepackage{soul}

\newcommand{\be}{\begin{equation}}
\newcommand{\ee}{\end{equation}} 

\newcommand{\avphi}{\langle\vert\phi\vert^2\rangle}

\title{Properties of the electroweak vacuum versus the QCD-vacuum in strong magnetic fields}

\ShortTitle{Electroweak versus QCD-vacuum in strong magnetic fields}

\author{Jos Van Doorsselaere\\
       Ghent University, Belgium\\
       E-mail: \email{jos.vandoorsselaere@ugent.be}}

\FullConference{International Workshop on QCD Green's Functions, Confinement and Phenomenology\\
		 5-9 September 2011\\
		 Trento, Italy}

\abstract{
	In strongly magnetized backgrounds, electrically charged vector bosons can condense. When considering the $\rho$--meson sector of QCD, this results in the existence of a superconducting state with Abrikosov vortices. As a next order effect, also the neutral mesons condense and one expects the resulting lattice to be a mix of neutral and charged condensates, opening a possibility to the existence of a superfluid property. We will show how the detailed structure of the vacuum can be calculated and which implications it has on the nature of the exotic superconducting state. 
	}

\begin{document}

\section{Introduction}

	It was proposed last year \cite{Chernodub:2010qx} that in a strong magnetic field the vacuum may become an electromagnetic superconductor. For very strong magnetic fields, the lightest of the charged vector mesons are likely to condense in a similar way as the electroweak vector bosons \cite{Ambjorn:1988tm}.  The $\rho$--mesons have a polarization with the right quantum numbers to see its energy reduced by an increasing magnetic field, a property that is not shared with the lighter $\pi$--mesons. Hence the validity of the restriction to the $\rho$-meson sector and the stability of the condensates, as $\pi$--mesons are no longer favored energetically.
	
	While it is a hope to observe the effects of such condensates experimentally, the required \emph{static} magnetic field is way beyond experimental capabilities. However, a beam collision with very small impact parameter could create a magnetic field with the required strength \cite{Skokov:2009qp}, about $10^{16}T$. This setting is promising, but does share the absence of electrical fields with the theoretical approximation. It is unclear whether the new vacuum could still appear in this setup and if it could create a detectable signal. If not, the relevance may be in Big Bang scenarios, as the early universe may have been strongly magnetized \cite{Grasso:2000wj} and have suffered some effects of the exotic state this implies, as we will show.
	From a theoretical point of view there's also a strong incentive to study this effect, as the results seem to have a broader application base than the $\rho$--mesons only. For example,  $W$-condensation in the electroweak model \cite{Ambjorn:1988tm}  can be  better understood by the very same --or at least similar-- calculations and in particular one finds that the condensates imply superconductivity \cite{ournextpaper}. The obtained solution to the equations of motion joins other well known, non-trivial such solutions like monopoles, cosmic strings and sphalerons, all interesting in their own right\cite{acchucarro}.
	
	Aside from obtaining a better picture of the vacuum configuration and its superconducting nature, the calculations presented give some insight in the potential superfluid structure of the vacuum. While no proof of superfluidity was found, and in particular the superfluid London equations are not satisfied, we find a substructure in the Abrikosov lattice consisting of 'superfluid' vortices \cite{Chernodub:2011}. These vortices give a non-trivial winding of the phase of neutral fields, just like in superfluids, but it remains unclear whether our exotic medium is a superconducting/superfluid hybrid or just a very peculiar superconductor. The superconductivity, at least, was confirmed in numerical lattice calculations \cite{Braguta:2011hq} and also hinted from holographic approaches \cite{Callebaut:2011ab}.
		
	As a last interesting fact, we found the superconductivity to be strongly influenced by the neutral condensate such that, in the core of the Abrikosov-vortices the London-equation is still satisfied, unlike the ordinary case where it vanishes at the core. This means in some sense the superconductivity is stronger than in the ordinary case. 
	
	In these proceedings we will go into detail on how the Abrikosov lattice can be obtained starting from nothing but the Lagrangian of the theory. It corresponds roughly to the paper \cite{Chernodub:2011}, work done in collaboration with Maxim Chernodub and Henri Verschelde.

\section{Effective potentials}

	To find an effective potential that can give a description of the Abrikosov lattice we will basically need three steps: we describe the model in the transversal (to the external magnetic field) plane, then make a small field approximation and eventually solve the simplified equations of motions and write the energy density as a function of one scalar field only. We will repeat the steps for three interesting cases: the ordinary Ginzburg-Landau model for comparison, the electroweak theory studied by Ambjorn and Olesen and the $\rho$--meson sector of the QCD-vacuum, described by the DSGS model \cite{Djukanovic:2005ag}. Many of the calculations can be found already in \cite{Chernodub:2010qx} and more recent \cite{Chernodub:2011}.
	
	\subsection{Ginzburg-Landau model}
	
		The Lagrangian we start from is given by 
			\begin{align}
			L&=\vert D_\mu \phi\vert^2-V(\vert\phi\vert^2)-\frac{1}{4}F_{\mu\nu}^2;\\ V(x)&=\frac{1}{4}\lambda (x-\phi_0^2)^2;\quad  D_\mu\phi=\partial_\mu\phi-ieA_{\mu}\phi.
			\end{align}
		With the identification
			\be
			\mathcal{O}=\mathcal{O}_1+i\mathcal{O}_2;\quad \mathcal{\bar O}=\mathcal{O}_1-i\mathcal{O}_2 ;
			\ee
		for any operator, we can rewrite this as an energy density, putting all longitudinal ($(0,3)$--directional) dependence to zero and lowering all (spacial) indices.
			\be
			\mathcal{E}= \vert \bar D\phi\vert^2+eF_{12}\vert\phi\vert^2+V(\vert\phi\vert^2)+\frac{1}{2}F_{12}^2\label{energy}.
			\ee
		Here surface terms are neglected and we used $\lbrack D,\bar D\rbrack\phi=-2eF_{12}\phi$. 
		
		Now we are ready to make an approximation near the critical point for a sufficiently high external field $eB=e\langle F_{12}\rangle =\frac{\lambda}{2}\phi_0^2$, where all terms in the above expression become positive.
			\begin{align}
			eF_{12}>\frac{\lambda}{2}\phi_0^2&\Rightarrow \phi=0\\
			eF_{12}\lesssim \frac{\lambda}{2}\phi_0^2&\Rightarrow \bar D\phi= \mathcal{O}(\phi^2)\sim 0
			\end{align}
		Neglecting the contribution of the covariant derivative in a well-chosen gauge as well as a periodicity requirement for the lattice, gives an Abrikosov ansatz \cite{Abrikosov} for test-space of solutions for $\phi$, that will be used for a variational method:
			\be 
			\bar D\phi_n=\bar \partial\phi_n-\frac{e}{2}B\bar z\phi_n=0\ \Rightarrow\  \phi_n=e^{-\frac{e}{4}B\vert z \vert^2}e^{-2\pi n \frac{\bar z}{L_B}-\frac{e}{4}B\bar z^2-\pi n^2},
			\ee
		with $L_B=\sqrt{\frac{2\pi}{eB}}$. It is easily seen that with this form translations in real and imaginary directions give ($n,N\in\mathbb{N}$):
			\begin{align}
			\phi_n(z+iL_B,\bar z-iL_B)&=e^{i\frac{z+\bar z}{2}}\phi_n(z,\bar z)\\
			\phi_n(z+NL_B,\bar z+NL_B)&=e^{i\pi\frac{iz-i\bar z}{2L_B}}\phi_{(n+N)}(z,\bar z)
			\end{align}
		Neglecting the $n$ independent gauge transformation, we find a periodic solution 
			\be 
			\phi(z,\bar z)=\sum_nC_n\phi_n(z,\bar z)\label{ansatz}
			\ee
		if $C_{n+N}=C_n$ for some $N$ and for all $n$. Fixing $N$, we now have an $N$--parameter ansatz to be plugged into the energy functional.
		
		Lastly we determine the appropriate energy from (\ref{energy}).  The first term can already be dropped, as our test function makes it negligible. The equation of motion for the gauge field gives furthermore:
			\be
			\partial (F_{12}+e\vert\phi\vert^2)=0
			\ee			
		which implies the argument is constant and equal to its average value, thus:
			\be
			F_{12}=B+e\langle\vert\phi^2\vert\rangle-e\vert\phi\vert^2.
			\ee
		The last step is straightforward:
			\be 
			\langle\mathcal{E}\rangle=(eB-\frac{\lambda}{2}\phi_0^2)\avphi+\frac{e^2}{2}\avphi^2+(\frac{\lambda}{4}-\frac{e^2}{2})\langle\vert\phi\vert^4\rangle+\phi_0^4
			\ee
		Plugging in (\ref{ansatz}) allows a numerical minimization in the $N$--dimensional parameter space. 
		
		Interestingly, the third term in the above expression can be rewritten in the vacuum solution as:
			\be
			\beta_A(\frac{\lambda}{4}-\frac{e^2}{2})\avphi^2
			\ee
		with $\beta_A$ the Abrikosov-parameter determining the lattice packing, unlike $\avphi$ that is independent of it. When $\beta_A=0$ this ordering becomes degenerate, indicating a phase transition. Indeed, this corresponds to the Bogomolnyi limit at which type II superconductivity goes to type I and vice versa.
		
		A last point we can look at is an explicit proof that we have superconductivity. Therefore we need the covariant form of the London-relation
			\be
			\frac{\partial J^i}{\partial t}=\frac{n_s e^2}{m} E^i
			\ee
		with $E$ the electrical field and $\sigma$ the conductivity, essentially depending on the density of cooper-pairs. We can determinge equations of motion assuming a very small electrical field in the longitudinal direction (parallel to the external magnetic field, but all other components still transversal. We get for the current $J_\mu=\partial^\nu F_{\nu\mu}$:
			\be
			\partial_0J^3-\partial^3J_0=2e^2\vert\phi\vert^2E^3\label{london}
			\ee

	\subsection{Electroweak vacuum}
		
		Now that we have gone through the simplest case in much detail, we can apply this to a more complicated model. The Lagrangian of the electroweak model in the unitary gauge reads:
			\be
			L=-\frac{1}{4}(W^a_{\mu\nu})^2-\frac{1}{4}X_{\mu\nu}^2+\vert D_\mu\phi\vert^2-V(\vert\phi\vert^2),\qquad D_\mu\phi=\partial_\mu\phi-i\frac{g}{2\cos\theta}Z_\mu\phi,
			\ee
		with the photon field  $A_\mu=\cos\theta X_\mu+\sin\theta W^3_\mu$, $Z_\mu=-\sin\theta X_\mu+\cos\theta W^3_\mu$ and $W^-_\mu=(W^1_\mu+iW^2_\mu)/\sqrt{2}$.
		
		Taking the convention $W=W^-_1+iW^-_2$, neglecting the component $\bar W$ and gauging to the unitary gauge, the expression for the energy density for transversal components only is:
			\begin{multline}
			\mathcal{E}=\frac{1}{2}F_{12}^2+\frac{1}{2}\vert \mathcal{\bar D}W\vert^2+\frac{1}{2}(\frac{g^2\phi^2}{2}-eF_{12}-g\cos\theta Z_{12})\vert W\vert^2+\frac{g^2}{8}\vert W\vert^4\\+\frac{1}{2}Z_{12}^2+\frac{g^2\phi^2}{4\cos^2\theta}\vert Z\vert^2+\vert  \partial\phi\vert^2+V(\vert\phi\vert^2)
			\end{multline} with $ \mathcal{\bar D}W=\bar \partial W+ie\bar A W+ig\cos\theta \bar Z W$.
		Knowing that in the ordinary vacuum $\phi$ gives a mass to the $W$, we observe the same compensating effect of the magnetic field in the mass term as we did in the ordinary superconductor. Very important is however that now the sign of this term is reversed, a direct consequence of the spin-1 nature of the $W$-field.
		It is easy to calculate that for positive $F_{12}$ this effect is absent for the $\bar W$ component and that it will not contribute a vacuum expectation value. Leaving it out was thus justified.
		
		Now we have
			\begin{align}
			eF_{12}<\frac{g^2}{2}\phi_0^2&\Rightarrow W=0\\
			eF_{12} \gtrsim\frac{g^2}{2}\phi_0^2&\Rightarrow \mathcal{\bar D}W= \mathcal{O}(W^2)\sim 0
			\end{align}
		We can in fact assume all terms to be the same small order $\mathcal{O}(W^4)$, as they are in an approximation up to $\mathcal{O}(W^2)$ all still vanishing.  With $Z\sim\phi-\phi_0\sim\mathcal{O}(W^2)$, we can solve their respective equations of motion and the one for the photon field and find:
			\be
			Z=i\frac{g}{2}\cos\theta\partial\frac{1}{\bar\partial\partial-M_Z^2}\vert W\vert^2
			;\quad
			\phi-\phi_0=\frac{g^2\phi_0}{4}\frac{1}{\bar\partial\partial-M_H^2}\vert W\vert^2; \quad F_{12}=B+\frac{e}{2}\vert W\vert^2-\frac{e}{2}\langle\vert W\vert^2\rangle;
			\ee
		with $M_Z^2=\frac{g^2\phi_0^2}{2\cos^2\theta}$ and $M_H^2=\lambda\phi_0^2$. 
		This is very similar to the Abelian case, but with a different sign for $\vert W\vert^2$, indicating the magnetic field and the superconducting condensate are in fact proportional to each-other, in stead of the usual inverse correlation. This property is called \emph{anti-ferromagnetism} and it indicates that the superconducting state, rather than being destroyed at a sufficiently high magnetic field, is created at a certain critical field strength.
		
		By assuming the very same ansatz (\ref{ansatz}) for $W$ as before, we can find a variational lattice solution if we rewrite the energy functional in terms of the $W$-condensate only. This involves some careful bookkeeping in coefficients, but is fairly straightforward:
			\be
			\langle\mathcal{E}\rangle=\frac{e^2}{8}\langle \vert W\vert^2\rangle ^2+\frac{1}{2}(M_W^2-eB_{ext})\langle \vert W\vert^2\rangle+\frac{1}{2}B_{ext}^2+\frac{g^2}{8} M_W^2\left\langle \vert W\vert^2\left(\frac{1}{\Delta-M_H^2}-\frac{1}{\Delta-M_Z^2}\right)\vert W\vert^2\right\rangle.
			\ee		
		Clearly, the Bogomolnyi limit appears for $M_Z=M_H$, exactly as was already shown in Ambjorn and Olesen's original work. 
		
	\subsection{QCD vacuum}
	
		Armed with insight from the previous models, we can put the machinery to work in case of an effective model for the $\rho$--meson sector of QCD. As argued in the introduction, this will hold the dominating contribution to a vacuum condensate in a strong magnetic field. The DSGS Lagrangian is given by \cite{Djukanovic:2005ag}:
			\be
			L=-\frac{1}{4}(\rho_{\mu\nu}^a)^2-\frac{1}{4}F_{\mu\nu}^2+\frac{m_\rho^2}{2}(\rho^{a}_\mu)^2+\frac{e}{2g_s}\rho^{(0)}_{\mu\nu}F^{\mu\nu}.
			\ee
		For consistency, we will assume the same notation for the charged mesons as we did previously for the vector bosons when rewriting the theory in the transverse plane. It is good to know that the convention in the original paper \cite{Chernodub:2010qx} is found by $\bar\rho\mapsto 2\rho$ in the formula below. 
		
		Neglecting the $\rho$-polarization in favor of the $\bar\rho$ for the same reason as we dropped $\bar W$ previously, but the other way around due to the conventions in \cite{Djukanovic:2005ag,Chernodub:2010qx}. We obtain for the transversal degrees of freedom:
			\be
			\mathcal{E}=\frac{1}{2}F_{12}^2+\frac{1}{2}\vert \mathcal{D}\bar \rho\vert^2+\frac{1}{2}\left(m_\rho^2+g_sf_{12}^{(0)}-eF_{12}\right)\vert\bar\rho\vert^2+\frac{g_s^2}{8}\vert\bar\rho\vert^4+\frac{1}{2}(f_{12}^{(0)})^2+\frac{1}{2}m_\rho^2\vert\rho^{(0)}\vert^2-\frac{e}{g_s}F_{12}f^{(0)}_{12}
			\ee
		with
			$
			\mathcal{D}\rho=\partial\bar\rho-ie A\bar\rho+ig_s\rho^{(0)}\bar\rho
			$ and $
			 f^{(0)}_{12}=-\frac{i}{2}(\bar\partial\rho^{(0)}-\partial\bar\rho^{(0)}).
			$
			The similarity with the electroweak model is obvious and we proceed in exactly the same way, making the appropriate estimate for all fields assuming $e\langle F_{12}\rangle=eB\gtrsim m_\rho^2$. The equations of motion give us
			\be
			\rho^{(0)}=-i\frac{g_s}{2}\partial\frac{1}{\bar\partial\partial-m_0^2}\vert\bar\rho\vert^2;\quad F_{12}=B+\frac{e}{2}\vert\bar\rho\vert^2-\frac{e}{2}\langle\vert\bar\rho\vert^2\rangle-\frac{e}{2}\bar\partial\partial\frac{1}{\bar\partial\partial-m_0^2}\vert\bar\rho\vert^2;
			\ee
		using $\langle f_{12}^{(0)}\rangle=0$ and $m_0^2=m_\rho^2/(1-\frac{e^2}{g_s^2})$. This leads tediously to
			\be
			\langle\mathcal{E}\rangle=\frac{1}{2}B_{ext}^2+\frac{e^2}{8}\langle\vert\bar\rho\vert^2\rangle^2+\frac{1}{2}(m_\rho^2-eB_{ext})\langle\vert\bar\rho\vert^2\rangle+\frac{g_s^2}{8}m_\rho^2\left\langle\vert\bar\rho\vert^2\frac{1}{-\partial\bar\partial+m_0^2}\vert\bar\rho\vert^2\right\rangle,
			\ee
		where we replace $\bar\rho$ by the Abrikosov ansatz illustrated before and proceed to the numerical minimization.
		
		It must be noted that the coefficient of the last term is much larger than in the electroweak case, as there's no compensating second term in it. This makes the algoritm much more stable, and is exactly the reason we obtained so far only results for this case, the electroweak structure should soon follow \cite{ournextpaper}.
		
	\section{Results}
	
		\begin{figure}[h]
		\begin{center}
		\begin{tabular}{ll}
		\includegraphics[height=40mm, angle=0]{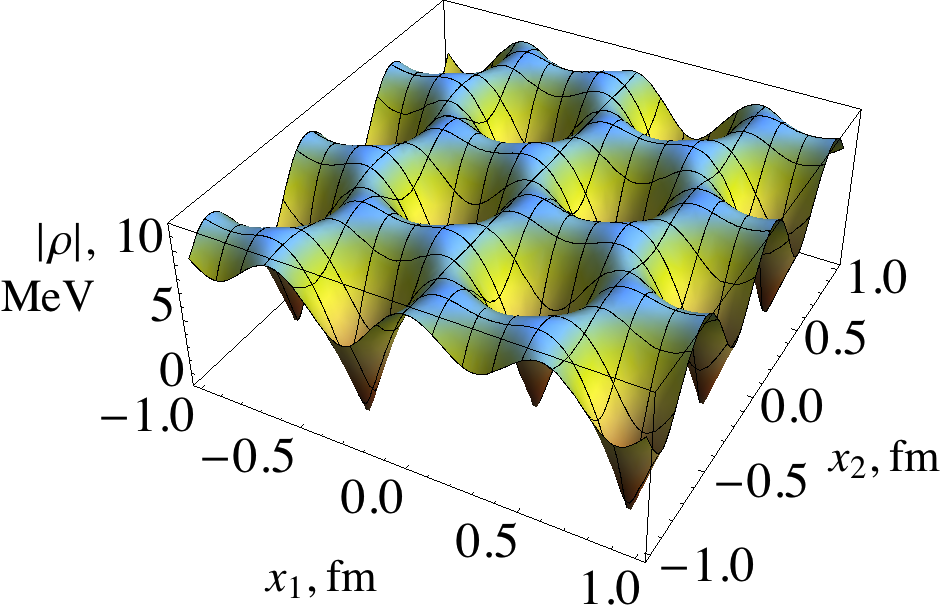} 
		& \hskip 5mm
		\includegraphics[height=40mm, angle=0]{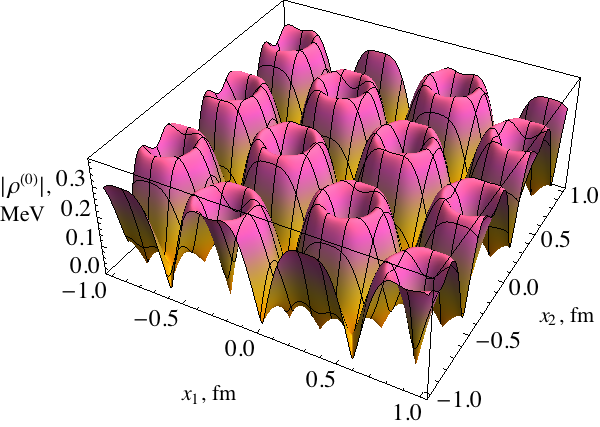}
		\end{tabular}
		\end{center}
		\caption{Structure of the charged (left) and neutral (right) condensates.}
		\label{fig:GL:lattices}
		\end{figure}

		The results show an ordering of vortices in a hexagonal structure, which can be obtained in a 2--parameter Abrikosov ansatz, reducing the energy of the more obvious 1--parameter square structure only slightly. More parameters do not improve the result. The $\bar\rho$-condensate lattice is superimposed with a more involved lattice of much smaller neutral vortex-like structures, that we call superfluid-vortices.

		The superconductivity is mediated by the charged condensate. While we now have proof it is organized hexagonally, the structure of the vortices is not essentially different from previous results \cite{Chernodub:2010qx, Ambjorn:1988tm}. What is interesting is the London relation, which becomes more involved by interaction with the neutral condensate. One gets:
			\be
			\partial_0J^3-\partial^3J_0= e^2 m_0^2 \Bigl( \frac{1}{- \bar\partial\partial + m_0^2}  |\bar \rho|^2\Bigr)E^3
			\ee
		compared to the result for the ordinary superconductor (\ref{london}) we have that the right hand side coefficient is much more 'smoothed out', and in particular does not make the superconductivity vanish at the vortex core.
		
		A second interesting discovery is the existence of the superfluid vortices: configurations with a non-trivial winding of the complex phase. The essential difference with the superconducting vortices is that the latter tell something about the phase of the charged fields, which is in this case a gauge parameter. This allows  for the winding property to create a topologically stable lattice. If the gauge parameter were not local, we would get a Goldstone boson and see a massless fluctuation, rather that seeing it eaten by a photon turned massive. This phenomenon is superfluidity, and can be seen in cold liquid Helium by the lack of both translational and rotational resistivity for low velocities.
		
		\begin{figure}[h]
		\begin{center}
		\includegraphics[height=40mm, angle=0]{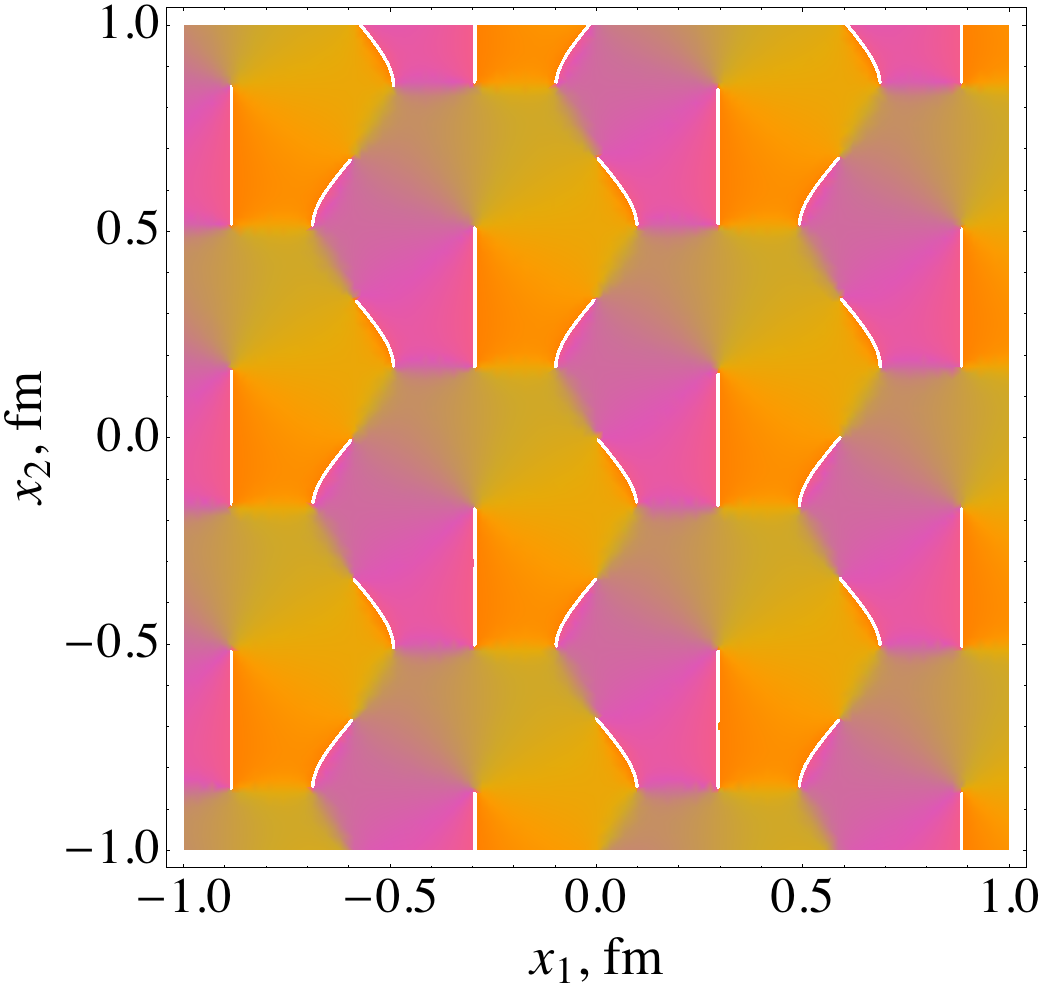} \hskip 15mm
		\includegraphics[height=40mm, angle=0]{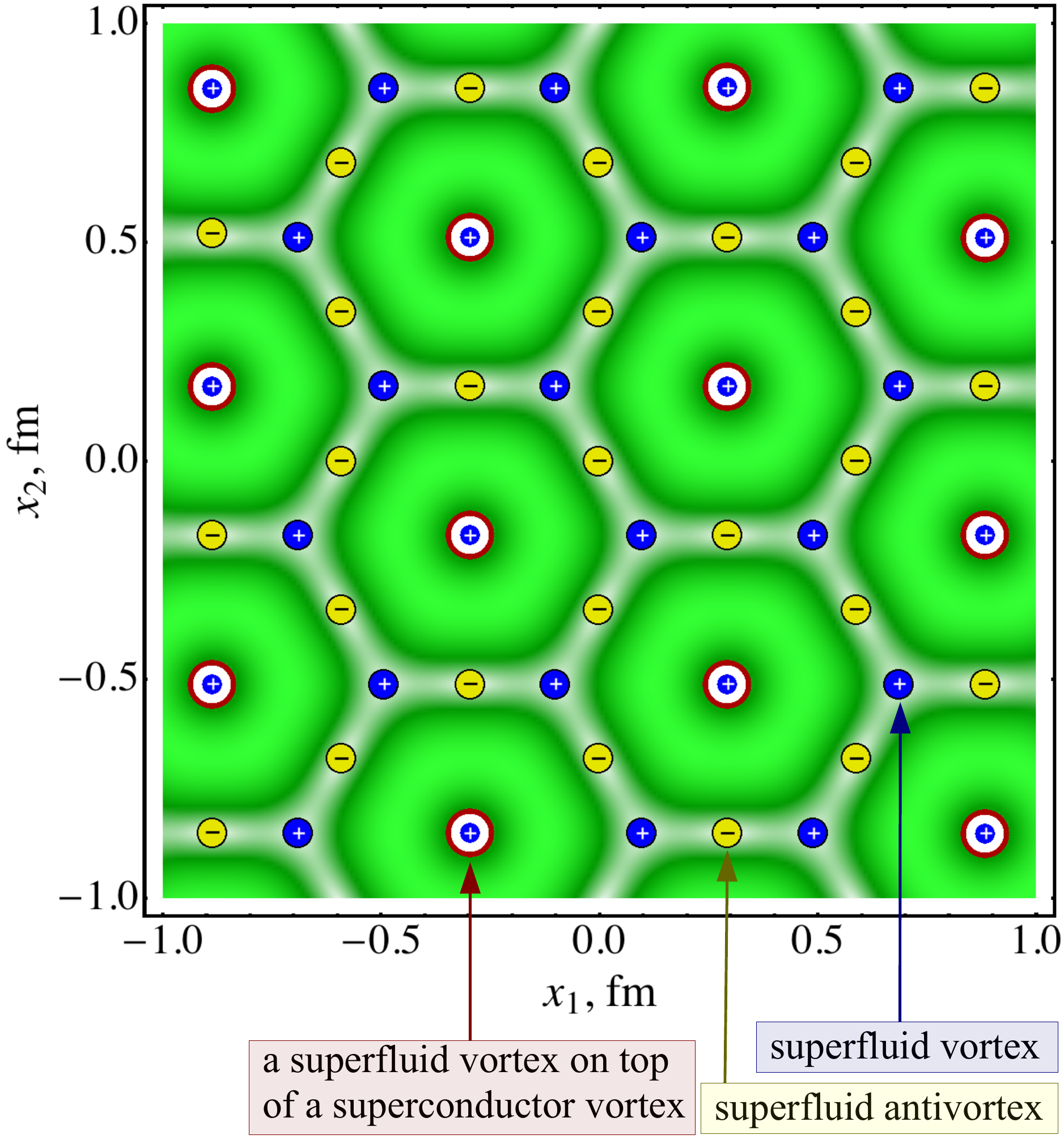}
		\end{center}
		\caption{Phase of the neutral condensate: left the argument explicitly plotted, right the vorticity of the branch points. Red circled are branch points to both $\rho^{(0)}$ and $\bar\rho$.}
		\end{figure}
		
		The lower order vortices in our superconducting background have indeed a non-trivial winding of a non-gauged phase of the neutral fields, hence we call them superfluid. But while the phase of $\rho^{(0)}$ is no symmetry of the DSGS Lagrangian, no massless states are present and there is no topological stability. Rather the superfluid vortices are stabilized by the superconducting vortices, and we could for example not have an isolated vortex solution. We can see from the plots clearly that the superfluid vortices come in pairs, and do not have a net vorticity.

\section{Outlook}
		
	While we have now computed the structure of the magnetized vacua close to the critical magnetization, an approximation for larger magnetic fields would be interesting. The equations are unfortunately much more involved once the condensates get larger. An interesting limit that may be in reach could be the case of magnetic field greatly exceeding all mass parameters in the model. For the electroweak model this means symmetry breaking is no longer caused by the Higgs field and a new phase appears. For the $\rho$--mesons no such effect exists, but as the masses become negligible, we could approach a phase with more symmetry. The hopes would be then to infer the intermediate regime from both near-critical solutions.
	
	Within the near-critical magnetization considered here, the existence of superfluidity is still an unsolved question. No straightforward evidence seems to follow from the theory, but further numerical research could shed light on the effect of translations and rotations of the lattice and possibly the existence of London-like equations.
	
	I  thank Daniele Binosi for the kind welcome at the TNT-II workshop and Maxim Chernodub and Henri Verschelde for collaboration on this subject.

\end{document}